\documentclass[fleqn,10pt]{wlscirep}
\usepackage[utf8]{inputenc}
\usepackage[T1]{fontenc}
\usepackage{ulem}
\usepackage{comment}
\usepackage{caption} % justify captions
\usepackage{xcolor}
\usepackage{colortbl}
\usepackage{float}
\usepackage{makecell}
\usepackage{multirow}
\usepackage{multicol}
\usepackage{mathtools}
\usepackage{amsmath}
\usepackage{amssymb}
\usepackage[small]{subfigure}
\usepackage{siunitx}
\usepackage{orcidlink}
\usepackage{pifont}

\definecolor{GREEN}{RGB}{0, 140, 80}

\makeatletter
\newcommand*{\starsection}[1]{%
  \section*{#1}%
  \NR@gettitle{#1}%
}
\makeatother
\makeatletter
\newcommand*{\starsubsection}[1]{%
  \subsection*{#1}%
  \NR@gettitle{#1}%
}
\makeatother
\makeatletter
\newcommand*{\starsubsubsection}[1]{%
  \subsubsection*{#1}%
  \NR@gettitle{#1}%
}
\makeatother

\newcommand{\figabvlabel}[1]{Fig.~\ref{#1}}
\newcommand{\seclabel}[1]{\nameref{#1}}
\newcommand{\tablabel}[1]{Table~\ref{#1}}

\definecolor{Wine}{RGB}{150,0,90}
\definecolor{DarkBlue}{RGB}{17,0,188}
\definecolor{Blue}{RGB}{0,19,243}
\definecolor{LightBlue}{RGB}{0,110,255}
\definecolor{Teal}{RGB}{24,209,197}
\definecolor{Green}{RGB}{92,255,83}
\definecolor{Lime}{RGB}{186,248,0}
\definecolor{Yellow}{RGB}{255,207,0}
\definecolor{Orange}{RGB}{255,99,0}
\definecolor{Red}{RGB}{255,0,0}

\title{Solving a Real-World Package Delivery Routing Problem Using Quantum Annealers}

\author[1,*]{Eneko Osaba\,\orcidlink{0000-0001-7863-9910}}
\author[1]{Esther Villar-Rodriguez\,\orcidlink{0000-0003-3343-3737}}
\author[2]{Antón Asla}

\affil[1]{TECNALIA, Basque Research and Technology Alliance (BRTA), 48160 Derio, Spain}
\affil[2]{Serikat - Consultoría y Servicios Tecnológicos, 48009 Bilbao, Spain.}

\affil[*]{eneko.osaba@tecnalia.com}

\keywords{Routing Problem, Package Delivery, Logistics, Optimization, Quantum Computing, Quantum Annealer, D-Wave}

\begin{abstract}

Research focused on the conjunction between quantum computing and routing problems has been very prolific in recent years. Most of the works revolve around classical problems such as the Traveling Salesman Problem or the Vehicle Routing Problem. The real-world applicability of these problems is dependent on the objectives and constraints considered. Anyway, it is undeniable that it is often difficult to translate complex requirements into these classical formulations.The main objective of this research is to present a solving scheme for dealing with realistic instances while maintaining all the characteristics and restrictions of the original real-world problem. Thus, a quantum-classical strategy has been developed, coined \texttt{Q4RPD}, that considers a set of real constraints such as a heterogeneous fleet of vehicles, priority deliveries, and capacities characterized by two values: weight and dimensions of the packages. \texttt{Q4RPD} resorts to the Leap Constrained Quadratic Model Hybrid Solver of D-Wave. To demonstrate the application of \texttt{Q4RPD}, an experimentation composed of six different instances has been conducted, aiming to serve as illustrative examples.

\end{abstract}
\begin{document}

\flushbottom
\maketitle

\thispagestyle{empty}

\starsection{Introduction}\label{sec:intro}

Routing problems are extensively studied in the fields of transportation and optimization \cite{kucukoglu2021electric,asghari2021green,konstantakopoulos2020vehicle}. The interest in this particular topic stems from two main factors: their high computational complexity (often categorized as NP-Hard \cite{hochba1997approximation}), which makes them challenging to address \cite{lenstra1981complexity}; and their broad applicability % logistic and business situations as well as tourism or leisure-related 
scenarios \cite{caceres2014rich}, entailing that groundbreaking improvements in the creation of effective routing algorithms have positive effects on society and industry.

For the computational complexity, it is crucial to emphasize that, for even relatively modest instances of routing problems, current computers struggle to run brute force methods. As a result, throughout the past few decades, numerous time-efficient solving schemes have been proposed, such as exact methods \cite{laporte1987exact}, heuristics \cite{rosenkrantz1977analysis}, and metaheuristics \cite{elshaer2020taxonomic}, with the last ones being the most commonly used. Furthermore, other interesting and advanced paradigms, such as Reinforcement Learning \cite{zhang2021solving} or Deep Learning \cite{xin2020step}, have also been explored to cope with challenging routing problems.

Most solvers developed so far have been designed to be run on conventional computing resources. Quantum computing (QC \cite{steane1998quantum}) has nevertheless become a promising alternative to these traditional devices for handling optimization and routing problems. Even though it is not a mature technology, QC has drawn much attention from the scientific community as it offers researchers and practitioners a revolutionary paradigm to solve different types of practical optimization problems \cite{gill2022quantum,yang2023survey}. Specifically, a large range of problems coming from domains such as healthcare \cite{ur2023quantum}, economics \cite{egger2020quantum}, industry \cite{luckow2021quantum}, energy \cite{ajagekar2019quantum}, or logistics \cite{osaba2022systematic} have benefited from QC lately.

The research on routing problems solved by QC approaches has been very prolific in recent years, with the Traveling Salesman Problem (TSP\cite{gutin2006traveling}) and Vehicle Routing Problem (VRP\cite{toth2002vehicle}) being the most studied cases. Specially interesting for this analysis is the extensive review published by Osaba et. al.\cite{osaba2022systematic}, stating that this new paradigm had motivated 53 research publications up to 2022. As concluded by the authors of that study, ``\textit{it is noticeable that the TSP engages most of the researchers (60,37\% - 32 out of 53 papers), while the VRP amounts to the 25,52\% of the contributions (13 out of 53). The rest of the papers deal with other routing problems, such as the Shortest Path Problem or the Hamiltonian Cycle}''. As of 2022, TSP \cite{delgado2022quantum,qian2023comparative,tszyunsi2023quantum,le2023quantum} and VRP \cite{mohanty2023analysis,poggel2023recommending,leonidas2023qubit,bentley2022quantum} keep maintaining a similar trend, accounting for the vast majority of scientific production.

The state-of-the-art might be classified according to their technical goals into two main branches. On the one hand, most of the works in the field pursue objectives such as unveiling the potential of quantum technologies or checking the efficiency of a particular method. This kind of work uses problems such as the TSP or VRP for benchmarking purposes. On the other hand, other studies present advances in the application of QC to real-world-oriented routing problems, analyzing aspects such as traffic congestion\cite{tambunan2022quantum}. This second category of research focuses on getting the most out of the current NISQ-era devices by implementing efficient and advanced hybrid resolution methods \cite{osaba2024hybrid}. As for the latter, the study of Weinberg et al. in 2023 \cite{weinberg2023supply} is probably the closest to solving a real routing problem using QC. The authors elaborate on the multi-truck vehicle routing for supply chain logistics. Being unable to solve the problem using a fully embedded approach, the authors propose an algorithm that iteratively assigns routes to trucks. This methodology allows the authors of that work to consider restrictions such as pick-up and drop-off demands and restricted driving windows.

In this context and aligned with this second category, this research proposes a novel solving scheme capable of providing a good, if not the best, solution to a real-world problem by making intelligent use of the capabilities offered by current quantum devices. To do so, the scheme, named \textit{Quantum for Real Package Delivery} (\texttt{Q4RPD}):

\begin{itemize}
    \item Focuses on solving a real-world routing problem defined by a Spanish company specializing in transport and logistics, named Ertransit\footnote{https://www.ertransit.com.cn/ertransit-espana/}. To the best of our knowledge, such a use case with delivery priorities, heterogeneous fleets, and two-dimensional descriptions of items has never been addressed in the literature from a quantum perspective. We, the authors, are confident that modeling realistic routing problems will push the limits of current research in QC and the advancement of the scientific community.
    \item Combines both quantum and classical computing in a hybrid method where:
    \begin{itemize}
         \item Classical computing controls the general workflow of \texttt{Q4RPD} in charge of splitting the problem into affordable sub-problems, providing the resources required to set up and do the computation, ensuring correct restrictions and preferences handling, and composing the final solution. All these steps are deeply described in Section~\seclabel{sec:fundamentals}.
        \item QC is applied to the calculation of each route, i.e., the trajectory associated with a truck, or a section of a route, via the Leap Constrained Quadratic Model (CQM) Hybrid Solver (\texttt{LeapCQMHybrid}\cite{leapCQM}) of D-Wave. This hybrid method solves problems formulated as CQM, which refers to a mathematical model defined by integer, real, and binary variables; linear, quadratic, inequality, and equality constraints; and a quadratic objective function\cite{leapCQM}. All these characteristics contribute to a more user-friendly and comprehensive framework in comparison to the native Quadratic Unconstrained Binary Optimization (QUBO) formulation of most of the Quantum Processing Units (QPUs).    
    \end{itemize}
    \item Demonstrates its application through an experimentation with six different instances aiming to serve as illustrative examples. Paraphrasing the words of Quetschlich et al. \cite{quetschlich2024utilizing}, ``\textit{the development of quantum computing applications at the moment considers mostly toy-size problem instances}". This same situation is highlighted by Weinberg et al. \cite{weinberg2023supply} who claim that ``\textit{there is a growing body of literature that tests quantum algorithms on miniaturized versions of problems that arise in an operations research setting}". In this regard, the iterative nature of \texttt{Q4RPD} has allowed the addressing of instances close to those handled by Ertransit in day-to-day operations, both in terms of complexity and size.
    
\end{itemize}

The rest of the article is organized as follows: in the following section, the~\seclabel{sec:problem} is introduced. After that, in~\seclabel{sec:fundamentals}, the characteristics of the method implemented are described. After that, in~\seclabel{sec:formulation}, the formulation of the problem addressed by the quantum device is presented. In~\seclabel{sec:results}, the applicability of the implemented system is demonstrated employing a set of instances as input. ~\seclabel{sec:conc} closes this paper by highlighting the main conclusions drawn from the results obtained in the experimentation and the planned future work.

\starsection{Problem Definition}\label{sec:problem}

The routing problem addressed in this research can be defined as follows: given a set of last-mile deliveries to complete within the day (whose fulfillment in the daily time-span is feasible), a depot, and a fleet of heterogeneous and available owned and rental trucks, the \textit{\textbf{main objective}} is to calculate a set of routes that satisfy all of the predetermined demands while minimizing the total costs. Exceptionally, a customer may have more than one order.

For this purpose, a \textbf{route} is defined as 
\begin{quote}
\textit{a trajectory associated with a single truck (driven by a designated single driver) that completes a group of deliveries and must start and end at the depot.}
\end{quote} 
    
Furthermore, in this work, the total cost of a solution is the total distance of the planned routes and the prices associated with the use of the truck. In relation to the latter, if the truck is owned by Ertransit the cost is 0. Otherwise, the use of rental vehicles incurs a cost for the service provided. 

Additionally, the problem is subject to the following \textit{\textbf{constraints}}: 

\begin{itemize}
    \item \textit{R1}: the capacity of each truck is measured in terms of weight (\textit{kgs}) and dimension ($cm^3$) which limits the load to be carried on each route. At the same time, the deliveries to be distributed are also defined by these parameters. We can say, thus, that the representation of the capacities is \textit{two-dimensional}. 
    \item \textit{R2}: some of the packages must be delivered within a certain time frame (priorities). These deliveries are labeled with a Top-Priority (\texttt{TP}) tag. These priorities are similar to the time windows frequently-used in other research works \cite{kallehauge2005vehicle}, where the lower limit of the window is 0, and the upper limit 
    \item \textit{R3}: the duration of a route must be less than the driver's working day (which is the same for all the available workers). 
\end{itemize}

In this research, and for the sake of simplicity, distance and time of travel share the same value. %More specifically, the Euclidean distance has been used for this purpose. 
Lastly, the following \textit{\textbf{business preferences}} are defined, drawn from past real-world situations faced by Ertransit that became, over the years, the operating protocol. Note that the preference order in the list relates to their relevance (if two preferences clash, the one with the higher ranking must prevail):

\begin{itemize}
    \item \textit{P1}: a vehicle can only complete one route, meaning that the trucks cannot be reused on that day, whether or not this decision has an impact on cost. 
    \item \textit{P2}: the use of owned trucks must be prioritized, even if this implies a higher overall cost in terms of distance. The motivation for this preference is twofold: on the one hand, to avoid the transactions associated with renting a vehicle, and on the other hand, to minimize the possibility of having an accident with a non-company truck.
    \item \textit{P3}: preferential treatment must be given to solutions that use fewer vehicles, even if this implies a higher overall cost in terms of the distance. The purpose of this preference is to have as many trucks as possible in the depot in order to handle potential incidents or unplanned extra-routine situations.  
\end{itemize}

In light of this, this problem has been named \textit{2-Dimensional and Heterogeneous Package Delivery with Priorities} (\texttt{2DH-PDP}).

\starsection{Solving scheme and Fundamentals}\label{sec:fundamentals}

Given the aforementioned constraints and preferences, the resolution method implements a route search for each truck, i.e., it selects for a given truck the packages to be delivered and their delivery order. To efficiently deal with the temporal constraint \textit{R2} of the priority items, the concept of sub-route is introduced. A sub-route is a section of a truck's path that will eventually belong to a complete regular route. Working with sub-routes allows us to adapt the maximum travel time depending on whether there is a priority constraint \textit{R2} to resolve or not, in which case the time limit is imposed by constraint \textit{R3}.

 Having said this, the different kinds of trajectories that may be calculated according to their origin and destination are the following:

\begin{figure}[t]
    \centering
    \includegraphics[width=0.75\linewidth]{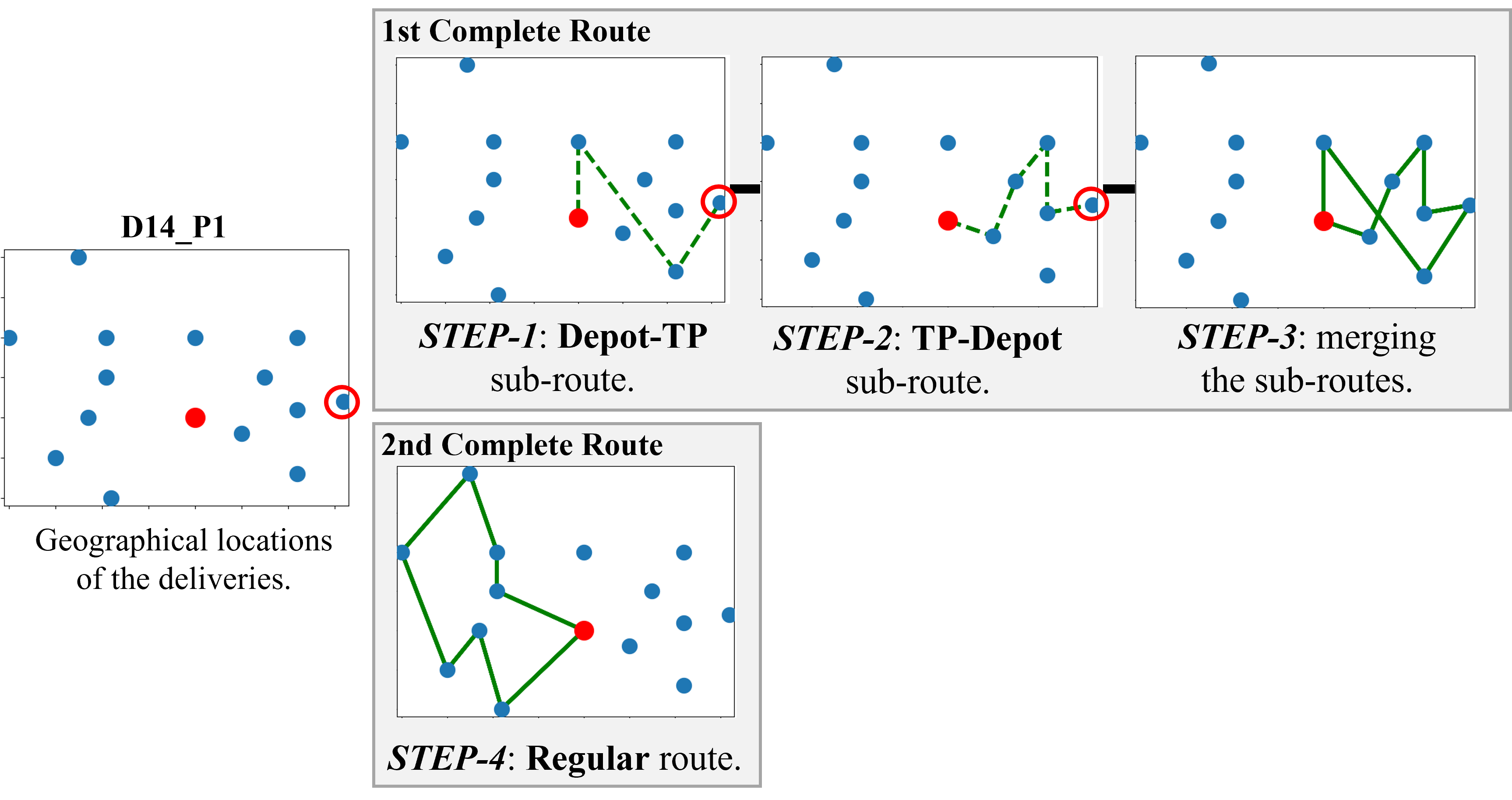}
    \caption{A graphical step-by-step resolution of an instance composed of 14 different deliveries (\texttt{D14\_P1}), being one of them a \texttt{TP} (surrounded by a red circle).}
    \label{fig:D15P1R2}
\end{figure}

\begin{itemize}
    \item \textit{Regular route}: this is a route completed by a single truck that starts and ends at the depot. This type of routes might be the result of a single route search process or of the chaining of several of the hereunder sub-routes. 
    Two \textit{regular routes} are depicted in STEP-3 and STEP-4 of Fig.\ref{fig:D15P1R2}.
    \item \textit{Depot-\texttt{TP} sub-route}: this is a sub-route that starts at the depot and finishes at the location of a \texttt{TP} delivery. The main requirement for this kind of sub-route is to arrive at the destination prior to the scheduled time set by the priority item. An example of a \textit{Depot-\texttt{TP}} sub-route is represented in STEP-1 of Fig.\ref{fig:D15P1R2}.
    \item \textit{\texttt{TP}-\texttt{TP} sub-route}: in this type of sub-route, both the origin and destination correspond to the locations of different \texttt{TP} deliveries. In this case, the time at which the subroute ends must be equal to or less than the time imposed by the priority package of the destination. 
    \item \textit{\texttt{TP}-Depot sub-route}: this is a sub-route which starts in a \texttt{TP} delivery location and ends in the depot. The STEP-2 of Fig.\ref{fig:D15P1R2} is a instance of this \textit{\texttt{TP}-Depot} sub-route.
\end{itemize}

\begin{figure}[H]
    \centering
    \includegraphics[width=0.85\linewidth]{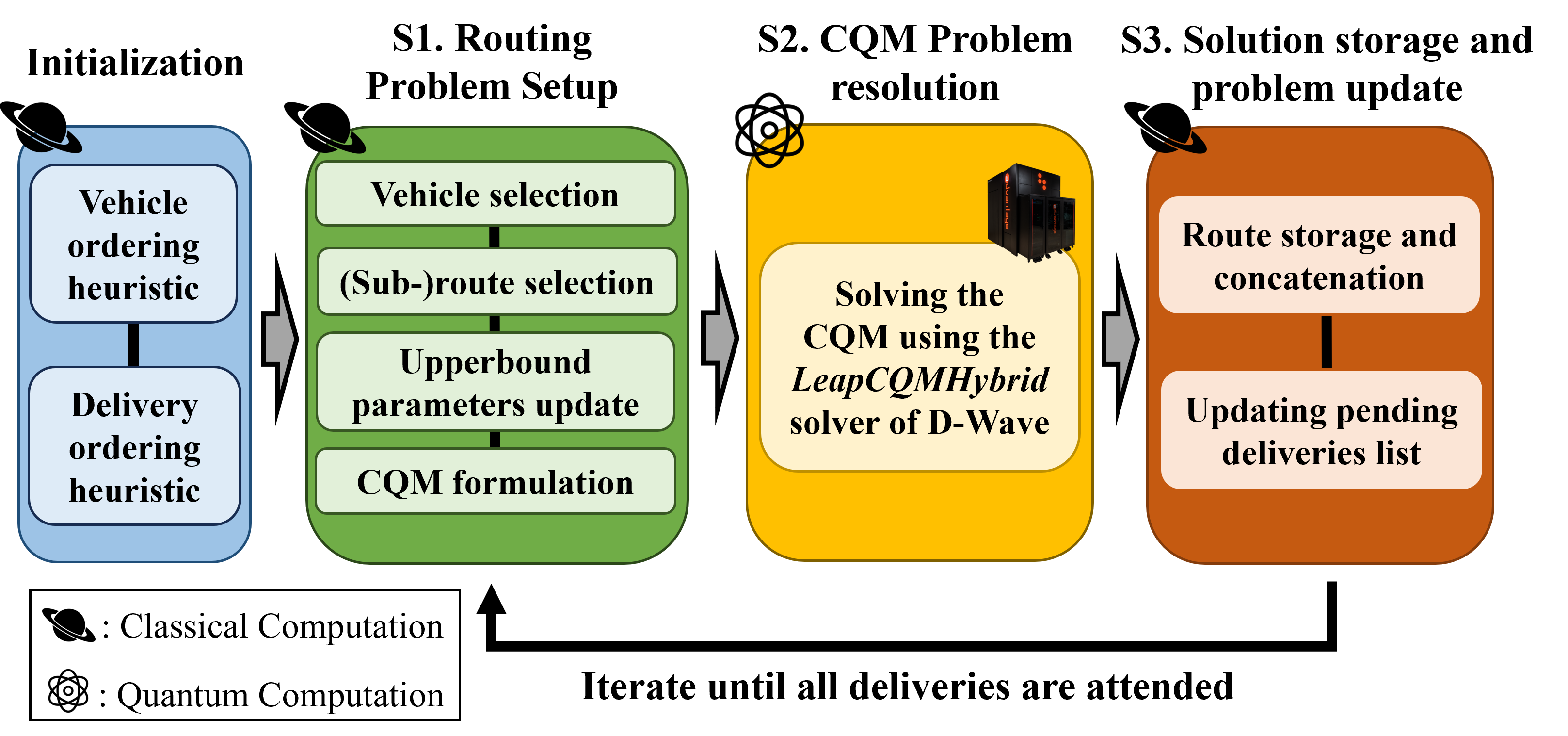}
    \caption{General workflow of \texttt{Q4RPD}.}
    \label{fig:Q4RPD}
\end{figure}

The \texttt{Q4RPD} solving scheme is based on an iterative process that, at each iteration, (1) picks out a truck and (2) calculates one trajectory (i.e. a regular route or a sub-route). Fig. \ref{fig:Q4RPD} represents the general workflow of \texttt{Q4RPD}. Right after loading the use case data but before starting the iterative process, \texttt{Q4RPD} begins by applying a pair of sorting heuristics:
\begin{itemize}
    \item \textbf{Vehicle ordering heuristic}: To satisfy preferences \textit{P2} and \textit{P3} introduced in~\seclabel{sec:problem}, vehicles are ordered and hence served by the following criterion: owned trucks, sorted by their global capacity, first; and rental trucks, ordered by their capacity, second. 
    \item \textbf{Delivery ordering heuristic}: the hard constraint \textit{R2} related to the deadline of the priority items, listed in the ~\seclabel{sec:problem} section is addressed by forcing the treatment of the restrictive deliveries first, i.e., when truck capacities and available times are at their maximum. When a \texttt{TP} delivery is dequeued, this item becomes the destination of a sub-route now limited by the timing restriction of this delivery. Note, therefore, that starting the \texttt{Q4RPD} by handling the \texttt{TP} deliveries does not mean that these packages will be the first to be delivered.
\end{itemize}

From this initialization onwards, an iterative solving scheme is in charge of partially solving the problem via a three-step procedure until the complete set of deliveries is satisfied:

\begin{enumerate}[label=\textbf{S\arabic*}.]
    \item \textbf{Routing problem setup}. This first step, which is executed using classical computing, is devoted to preparing the problem instance for being subsequently solved by the quantum device. This setup consists of the following methods: 
        \begin{itemize}
            \item \textbf{Vehicle selection}. \texttt{Q4RPD} picks the first truck in the list of available vehicles to calculate its most optimized route. It is important to clarify that the vehicle list definitively removes a truck only if it has been assigned a regular route, i.e., from depot to depot. This entails that if there is a truck whose route has not been completed yet, it will always be the first one on the list, thus prioritizing the completion of half-done routes over the start of a new one. At the same time, this procedure contributes to the rule stated by preference \textit{P3}.
            
            \item \textbf{(Sub-)route selection}: At this point, four different scenarios, summarized in Fig.\ref{fig:routeSelection}, may arise:
            \begin{enumerate} [label=(\Alph*)]
                \item If the selected truck had not completed its route, i.e., it was located at the \texttt{TP} point of a previously computed sub-route, and there are still prioritized deliveries to serve, \texttt{Q4RPD} checks if further \texttt{TP} deliveries are still \textit{reachable} by this vehicle. A \texttt{TP} delivery is \textbf{\textit{reachable}} if:
                \begin{itemize}
                    \item The active truck`s capacity is enough to serve the new demand, i.e., \textit{R1} is satisfied.
                    \item And the active truck can arrive on time to both this intermediate destination and the depot, and the driver´s working day will not be exceeded when completing the whole route, i.e., the \textit{R2} and \textit{R3} constraints remain satisfiable after accepting this new sub-route.
                \end{itemize}
                If a \texttt{TP} delivery were \textbf{\textit{reachable}}, a \textit{\texttt{TP}-\texttt{TP}} sub-route would be calculated. Otherwise, a \textit{\texttt{TP}-Depot} trajectory  would be the objective to complete its route (\textit{B} in Fig.\ref{fig:routeSelection}).
                \item If the selected truck had not completed its route but there were no \texttt{TP} items yet to be served, a \textit{\texttt{TP}-Depot} sub-route to complete the route would be built.
                \item If the truck was in the depot and there were still \texttt{TP} deliveries to serve, \texttt{Q4RPD} would take the first element out of the prioritized deliveries queue to calculate a \textit{Depot-\texttt{TP}} sub-route.
                \item If the truck was in the depot and there were no \texttt{TP} deliveries to serve, a \textit{regular} route would be calculated.
            \end{enumerate} 
            It should be noted that, in the end, each complete route would be a regular and cyclic path composed of either a single \textit{regular} route or a set of sub-routes.

             \begin{figure}[t]
                \centering
                \includegraphics[width=0.85\linewidth]{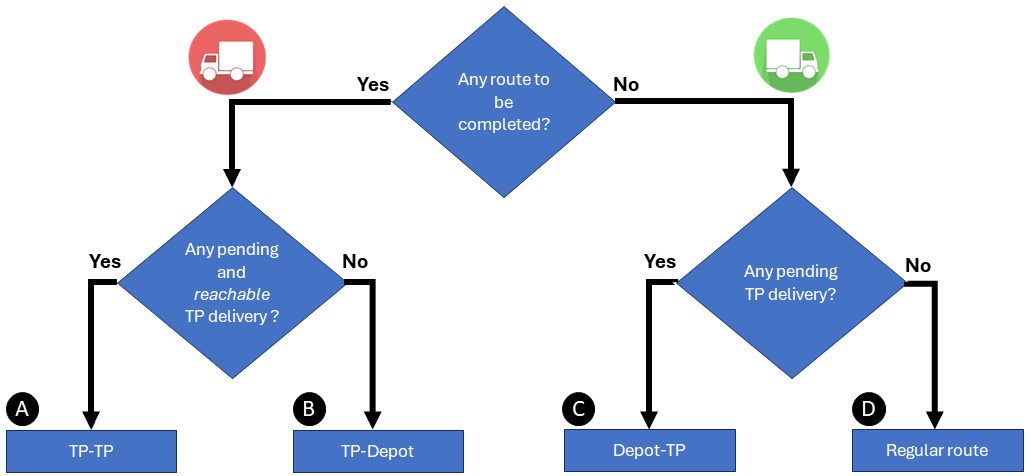}
                \caption{(Sub-)route selection procedure with the 4 scenarios.}
                \label{fig:routeSelection}
            \end{figure}
            
            \item \textbf{Upper bound parameters update}: the upper bound parameters (summarized in Table \ref{tab:upperbound_params}) cannot be fixed in the problem initialization but must be set at each iteration according to the type of the trajectory.
            
            \begin{itemize}
                \item \textbf{Maximum duration assignment}:
                the $rt$ of the new (sub-)route is calculated, embracing this criterion:
                \begin{enumerate}[label=(\Alph*)]
                    \item For a \textit{\texttt{TP}-\texttt{TP}} sub-route, with \textit{reachability} already confirmed, $rt$ is calculated as the difference between the destination's \texttt{TP} scheduled time and the time already spent since the beginning of the route, i.e., from the depot.
                    \item For a \textit{\texttt{TP}-Depot} route, $rt$ is the difference between the driver's working day and the time already spent since the beginning of the route.
                    \item For a \textit{Depot-\texttt{TP}} route, $rt$ is equal to the \texttt{TP} scheduled time.
                    \item For a \textit{regular} route, $rt$ is equal to the driver's working day.
                \end{enumerate} 
            \item \textbf{Usable truck capacity}: if the route of the active truck is still to be completed, capacities $W$ and $D$ must be updated by deducting the load already assigned to the truck in previous sub-routes from the vehicle's tare and dimension specification.
            \end{itemize}

            \noindent\begin{table}[t]
            \centering
            \begin{tabular}{p{2cm}p{9cm}}
            \hline
            {\bf Upper bound parameters} \\            
            $rt$ & the maximum possible duration of a (sub-)route. \\
            $W,D$ & the permissible maximum weight and dimension in the truck assigned to the (sub-)route. \\\hline
            \end{tabular}
            \caption{Upper bound parameters used in the formulation.}\label{tab:upperbound_params}
            \end{table}%

\item \textbf{CQM problem formulation}: since the employed \texttt{LeapCQMHybrid} library solves problems modeled as CQM, the last step of this stage is to build a CQM following the formulation described in the next section (\seclabel{sec:formulation}).
        \end{itemize}
\item \textbf{CQM problem resolution}: after the CQM is fomulated, this second step consists of the resolution of this CQM by the D-Wave's \texttt{LeapCQMHybrid} solver, which resorts to a quantum computer to process the query. 
As pointed out before, this CQM could model any of the routes or sub-routes above defined.
    \item \textbf{Partial solutions storage}: this last step, executed using classical computing, is devoted to updating the whole problem. Two procedures engage in this phase:
    \begin{itemize}
        \item Route storage and concatenation. The goal of this step is to store the complete routes of every truck. If the trajectory that has just been solved corresponds to a sub-route, it is appended to the previous sub-route of the truck, if any. In these cases, through concatenation, a regular route is finally composed once the destination of the trajectory is the depot. Accordingly, the truck with a complete route will be removed from the vehicle list.
        \item Update the list of pending deliveries, removing all the packages already dropped in the last solved (sub-)route. 
    \end{itemize}  
\end{enumerate}

\starsection{Single Routing Problem: Mathematical formulation}\label{sec:formulation}

This section introduces the mathematical formulation of the problem addressed by the \texttt{LeapCQMHybrid}, defined after the routing problem setup where the type of route to calculate and the vehicle were determined by the \textit{Vehicle ordering heuristic} (Step \textit{S1} in Fig.\ref{fig:Q4RPD}). As explained before and accordingly, the problem faced by the quantum device at each iteration is not the complete \texttt{2DH-PDP}, but the calculation of a single (sub-)route of any of the types listed in~\seclabel{sec:fundamentals}. For this reason, this sub-problem has been coined the \textit{Single Routing Problem} (\texttt{SRP}), whose variables and parameters are displayed in~\tablabel{tab:params_vars}. 
\noindent\begin{table}[!h]
\centering
\begin{tabular}{lp{14.2cm}}
\hline
\makecell[l]{\cellcolor{gray!10}\textbf{Upper bound} \\\cellcolor{gray!10}\textbf{parameters}} & \cellcolor{gray!10} \\ 
\cellcolor{gray!10}$rt$ & \cellcolor{gray!10}the maximum possible duration of a (sub-)route. \\
\cellcolor{gray!10}$W,D$ & \cellcolor{gray!10}the permissible maximum weight and dimension in the truck assigned to the (sub-)route. \\[2mm]
\makecell[l]{\textbf{Static} \\ \textbf{parameters}}\\
P & set of deliveries still to be met.\\
$M$ & number of deliveries (cardinality of P). \\
$w_i,d_i$ & weight and dimension of delivery $i\in$ P. \\
$t_i$ & the time limit for delivery $i$. \\
$c_{i,j}$, $d_{i,j}$, & travel time and distance associated with going from the location of delivery to the location of delivery $j$. \\[2mm]
\multicolumn{1}{l}{\bf Variables} \\
$x_{i,p}$ & binary variable that represents if the location of delivery $i$ is visited at position $p$ of the (sub-)route, with $p \in [0,M]$. \\ \hline
\end{tabular}
\caption{The complete set of parameters and variables used in our formulation.}\label{tab:params_vars}
\end{table}%

\starsubsection{Codification and variables}\label{sec:codification}

\begin{figure}[t]
    \centering
    \includegraphics[width=0.5\linewidth]{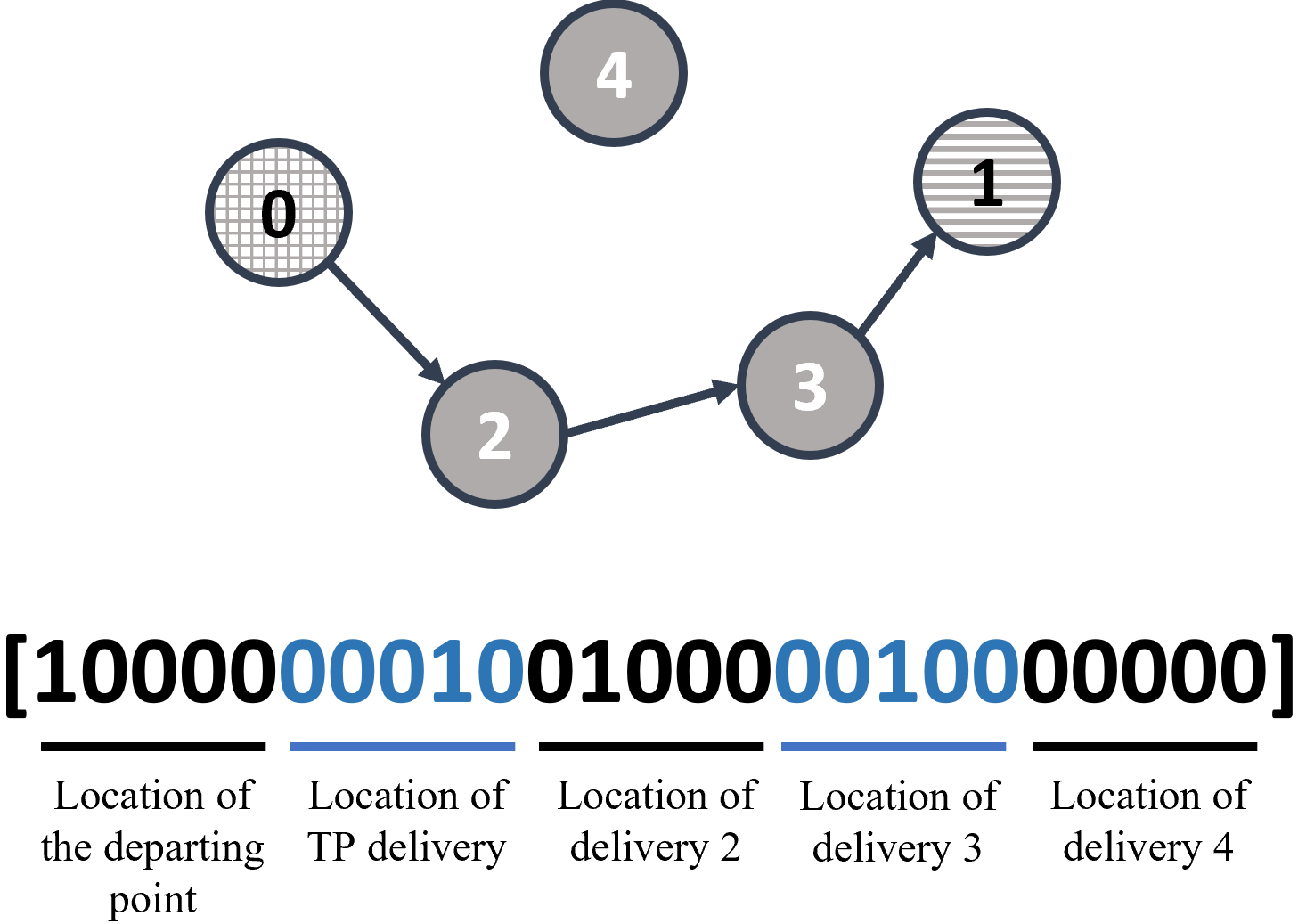}
    \caption{A \texttt{SRP} instance composed of five delivery locations, and a tentative solution. Delivery location with $ID$=0 represents the starting point of the route. Delivery location with $ID$=1 represents the destination, that is, the delivery location that must be reached before a certain restricted time.}
    \label{fig:5node}
\end{figure}

The binary codification used for the \texttt{SRP} is the one known as \textit{node-based}, often used in studies of quantum computing applied to routing problems \cite{weinberg2023supply,feld2019hybrid,salehi2022unconstrained}. A solution to the \texttt{SRP} is represented as a set $X=\{X_0,\dots,X_M\}$ of lists, in which each $X_i$ is associated with a single delivery $i$, where $M$ is the total number of deliveries and \textit{0} represents the origin location. Furthermore, $X_i=\{x_{i,0},\dots,x_{i,M}\}$, where $x_{i,p}$ is a binary variable used for representing the position of the delivery $i$ along the trajectory. Thus, $x_{i,p}$ is 1 if the location of the delivery $i$ is visited in position $p$ of the route and 0 otherwise. A visual example comprising five delivery locations is represented in Fig.\ref{fig:5node}.

So, the challenge when solving \texttt{SRP} is to find the most appropriate values for variables in $X$ in order to build the route that best optimizes the established objectives (later described in \seclabel{sec:obj}).

\starsubsection{Problem initialization}\label{sec:initialization}

As part of the problem modeling, a slight initialization procedure is carried out, which is comprised of two steps:

 \begin{itemize}
     \item Firstly, the two first $X$ components, i.e., $X_0$ and $X_1$ are reserved for the preset origin and destination of the (sub-)route, respectively. 
     \item Secondly, $x_{0,0}$ is set to 1 to ensure that the sub-route begins at the origin. In contrast, $X_1$, i.e., the set of variables identifying the position of the destination in the (sub-)route, cannot be fixed in advance since, at this stage, we do not know how many deliveries will eventually be included in the (sub-)route and, therefore, what the last position of the route will be.
 \end{itemize}

\starsubsection{Objectives}\label{sec:obj}

The \texttt{SRP} is an optimization problem that must be solved by minimizing a cost function defined by two different objectives. By multiplying each of these objectives by an appropriate weight, the user can determine the relevance given to each of them. Hence, $\min\text{ }\sum_{i=1}^2\omega_io_i$ represents the cost function, being $\omega_i$ the weight given to objective $o_i$. For this specific research, and as a result of an empirical study carried out in a laboratory environment, $\omega_1$=1 and $\omega_2$=2 have been applied. Therefore:

\begin{itemize}
    \item $o_1$: minimizing the total distance of the route, which can be mathematically formulated as follows:
        \begin{equation}\label{eq:o_2}
            o_1 = \min \sum_{i=0}^{M}\sum_{j=0}^{M}d_{i,j}x_{i,p}x_{j,p+1}\quad\forall p\in \{0,\dots,M-1\}, i\neq j.
        \end{equation}
    \item $o_2$: visiting the destination location as later as possible in the route, that is, maximizing the number of deliveries carried out before reaching the final destination. This objective is materialized by maximizing the value of $p$ in $x_{1,p}=1$. This objective is formulated as follows:
        \begin{equation}\label{eq:o_1}
            o_2 = \min \sum_{p=0}^{M}(-x_{1,p}-\sum_{p'=p}^{M}x_{1,p'}).
        \end{equation}
\end{itemize}

Finally, certain restrictions apply to the aforementioned objectives, which are essential to solving a real-world-oriented problem. The next subsection defines the entire pool of constraints (\seclabel{sec:constraints}).

\starsubsection{Problem constraints}\label{sec:constraints}
The above-described objectives are subjected to a set of seven different restrictions, which are:

\begin{itemize}
    \item \textbf{Delivery-consistency}: a delivery may not be on the route or, if it is, at most once.
    \begin{equation}\label{eq:consistency}
    \sum_{p=0}^{M}x_{i,p}\leq1\quad\forall i\in \{0,\dots,M\}.%, i\neq p.
    \end{equation}%
    This constraint differs from the classical formulations of the TSP \cite{matai2010traveling}, VRP \cite{toth2014vehicle}, and related problems in the sense that a route does not necessarily have to serve all deliveries.
    
    \item \textbf{Location-consistency}: each time slot of a truck schedule can be assigned to a single order at most. 
    \begin{equation}\label{eq:location}
    \sum_{i=0}^{M}x_{i,p}\leq1\quad\forall p\in \{0,\dots,M\}. %, i\neq p.
    \end{equation}%

    \item \textbf{Delivery-consecutiveness}: a correct calculation of objectives $o_1$ and $o_2$ requires that the deliveries served in the (sub-) route be consecutive, which means that there must be as many consecutive zeros from right to left in each $X_0, X_1,...X_M$ as there are surplus positions $p$.
    \begin{equation}\label{eq:consecutiveness}
    \sum_{i=0}^{M}x_{i,p}-\sum_{i=0}^{M}x_{i,p+1}\geq0\quad\forall p\in \{0,\dots,M-1\}. %, i\neq p.
    \end{equation}%

    \item \textbf{Destination-inclusion}: the destination must be in the route.
    \begin{equation}\label{eq:mostrestrictive}
    \sum_{p=0}^{M}x_{1,p}=1.
    \end{equation}%
    \item \textbf{Time-restriction}: the duration of the route must not be longer than $rt$, that is, the maximum possible duration of the route. 
    \begin{equation}\label{eq:time}
    \sum_{i=0}^{M}\sum_{j=0}^{M}c_{i,j}x_{i,p}x_{j,p+1} \leq rt \quad\forall p\in \{0,\dots,M-1\}, i\neq j.
    \end{equation}
    Note that, as aforementioned, in practice $c_{i,j}=d_{i,j}$.
    
    \item \textbf{Weight-restriction}: the maximum allowed weight $W$ for the truck must not be surpassed. 
    \begin{equation}\label{eq:weight}
    \sum_{p=0}^{M}x_{i,p}w_i \leq W \quad\forall i\in \{0,\dots,M\}.% i\neq p.
    \end{equation}
    \item \textbf{Dimension-restriction}: the maximum permitted dimension $D$ for the route must not be exceeded.  
    \begin{equation}\label{eq:dimension}
    \sum_{p=0}^{M}x_{i,p}d_i \leq D \quad\forall i\in \{0,\dots,M\}.%, i\neq p.
    \end{equation}
\end{itemize}

\starsection{Experimental results}\label{sec:results}

This section is devoted to analyzing the performance of \texttt{Q4RPD}. First, in \seclabel{sec:qs}, some details on the \texttt{LeapCQMHybrid} are provided. After that, in \seclabel{sec:bm}, the data employed in the experimentation is introduced. This section finishes in \seclabel{sec:pa} by showing the obtained results and analyzing the performance of \texttt{Q4RPD}.

\starsubsection{Quantum solver details}\label{sec:qs}
The hybrid solver used in this paper is the CQM model of \texttt{LeapCQMHybrid}, implemented by D-Wave System. In a nutshell, this method is part of D-Wave’s \textit{hybrid solver service} (HSS \cite{HSS}), which can be described as a portfolio of hybrid solvers implemented by D-Wave Systems. Methods included in HSS balance both classical and quantum computation to solve problems not fitting in QPUs. At the time this paper is being written, HSS accommodates three techniques to deal with three problem types: Binary Quadratic Models, Discrete Quadratic Models, and, lastly, CQMs.

\begin{figure}[t]
    \centering
    \includegraphics[width=0.45\linewidth]{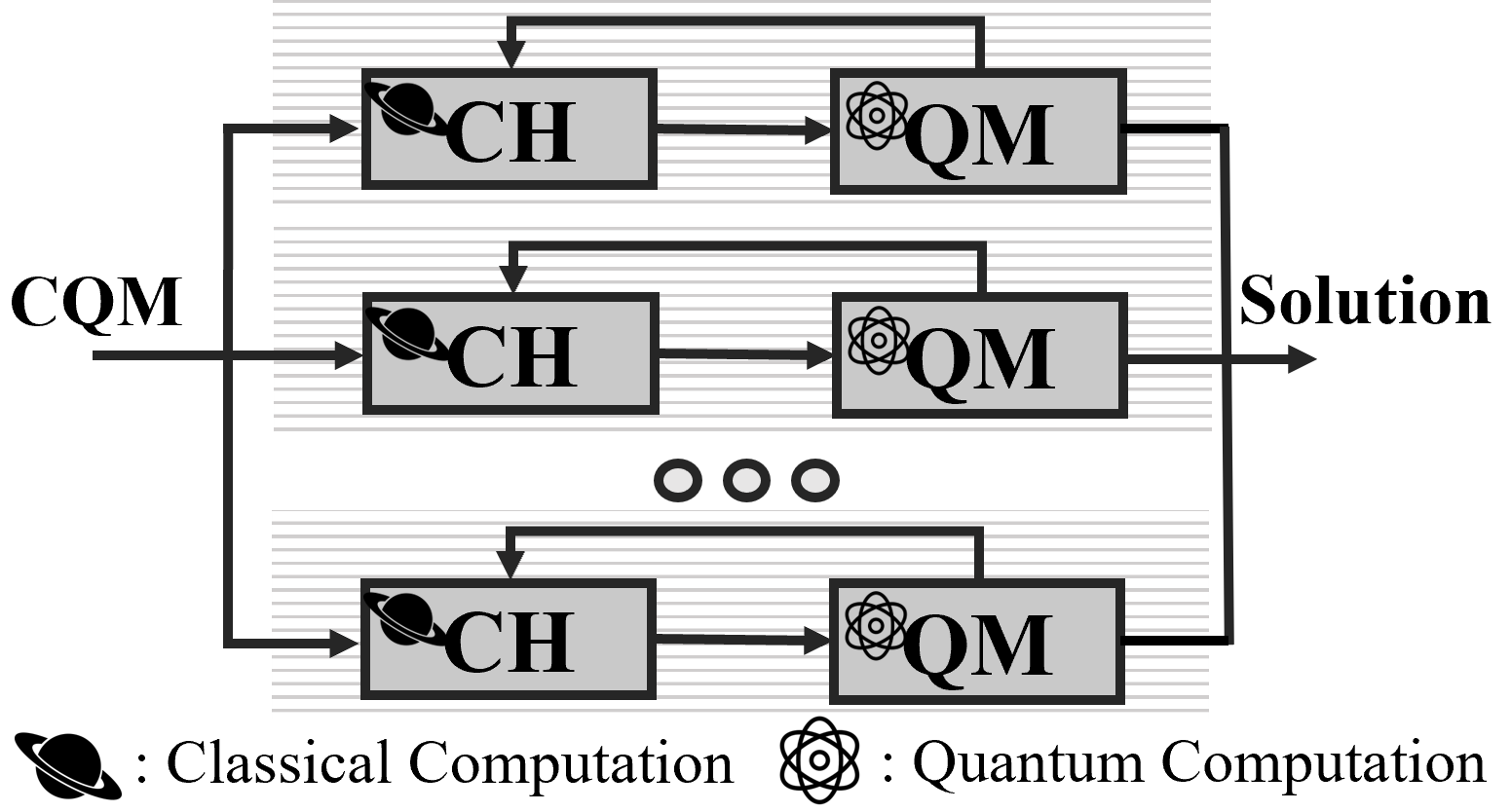}
    \caption{General scheme of \texttt{LeapCQMHybrid} solver. CH = Classical Heuristic Module. QM = Quantum Module.}
    \label{fig:LeapCQM}
\end{figure}

The \texttt{LeapCQMHybrid} workflow, depicted in Fig. \ref{fig:LeapCQM}, applies parallel processing to the search for solutions to eventually return the best solution found among the pool of threads. Each thread consists of:
\begin{itemize}
    \item a classical heuristic module (CH) with cutting-edge heuristic methods primarily aimed at solving the problem.
    \item and a quantum module (QM) in charge of guiding HM into promising areas of the search space and improving already-existing solutions by means of \textit{quantum queries}.
\end{itemize} 

In this research, the \textit{quantum queries} have been run on the \texttt{Advantage\_system6.4} architecture, which is the most recent at the time that this work was written. This computer features 5616 qubits, arranged in a Pegasus topology. Regarding the parameterization of the  \texttt{LeapCQMHybrid} solver, the only configurable parameter \texttt{time\_limit} was set to the default value for each instance. 

The \texttt{LeapCQMHybrid} solver is proprietary, therefore, further information about the quantum subroutines or the number of qubits required to face an instance of \texttt{SRP} is not available to the general public. For additional details on this method, we refer interested readers to the D-Wave report \cite{HybridDwave} and to recently published works focused on real-world applications of the \texttt{LeapCQMHybrid} \cite{glos2023optimizing,bozejko2023solving,bozejko2024optimal}.

\noindent\begin{table}[!t]
 \centering
 \resizebox{0.40\columnwidth}{!}{
     \begin{tabular}{|l|r|r|r|}
      \hline       
       \multicolumn{1}{|c|}{\textbf{Instance}} & \multicolumn{1}{c|}{\makecell[r]{Fleet \\Description}} & \multicolumn{1}{c|}{\#Variables} & \multicolumn{1}{c|}{\#Constraints}\\ \hline
      \texttt{D14\_P1} & 2\textit{o},3\textit{r} & 420 & 111 \\ \hline
      \texttt{D16\_P1} & 0\textit{o},4\textit{r} & 631 & 157 \\ \hline
      \texttt{D14\_P2} & 2\textit{o},3\textit{r} & 398 & 124\\ \hline
      \texttt{D21\_P2} & 3\textit{o},2\textit{r} & 834 & 209 \\ \hline
      \texttt{D21\_P0} & 2\textit{o},2\textit{r} & 702 & 138 \\ \hline
      \texttt{D29\_P0} & 0\textit{o},4\textit{r} & 1602 & 232 \\ \hline
     \end{tabular}
}
\caption{Characteristics of the six instances. The fleet is represented using X\textit{o},Y\textit{r} format, where X and Y are the number of \textit{o}wned and \textit{r}ental trucks, respectively.}\label{tab:instances}
\end{table}%

\starsubsection{Benchmark description}\label{sec:bm}
The benchmark of \texttt{Q4RPD} is composed of six different instances whose characteristics are summarized in Table \ref{tab:instances}. Because the problem addressed in this paper has been designed ad hoc for this research, it is not possible to find standard datasets to work with. This is why the six use cases employed have been generated specifically for this study, under the guidance of Ertransit.

Each instance is named \texttt{DX\_PY}, where \texttt{X} represents the size of the problem in terms of the number of deliveries and \texttt{Y} the amount of \texttt{TP} deliveries. The following information per instance is also provided in Table \ref{tab:instances}: the number of owned and rental trucks of the fleet of vehicles; and the number of variables and constraints involved in the global problem formulation.

The characteristics of each instance and the main motivations for their selection for benchmarking purposes are as follows:
\begin{itemize}
    \item \texttt{D14\_P1}: this is the smallest scenario which is composed of 13 non-priority and one priority delivery. The solution to \texttt{D14\_P1} can be seen in \figabvlabel{fig:D15P1R2}.
    
    \item \texttt{D16\_P1}: the main difference between this case and the previous one is the use of a drastically reduced working day (90 minutes) to push \texttt{Q4RPD} into a narrower feasible solution space since in \texttt{D14\_P1} the length of the routes is mostly restricted by the vehicle's capacity. The resolution process of this instance is shown in \figabvlabel{fig:D17P1R3}.
    
    \item \texttt{D14\_P2} and \texttt{D21\_P2}: the main characteristic of these scenarios is the existence of two priority deliveries that are not \textit{reachable} on the same route. In addition, \texttt{D14\_P2} has two orders from the same customer, which must be served by two distinct trucks to satisfy the restrictions while minimizing the objective function. \texttt{D21\_P2}, in turn, is the most demanding instance in terms of quantum resources employed. The solutions to \texttt{D14\_P2} and \texttt{D21\_P2} are depicted in \figabvlabel{fig:D15P2R2} and \figabvlabel{fig:D22P2R3}, respectively.
    
    \item \texttt{D21\_P0} and \texttt{D29\_P0}: these instances are characterized by being fully composed of non-priority deliveries, and their solutions are depicted in \figabvlabel{fig:D22P0R3} and \figabvlabel{fig:D30P0R4}. Regarding \texttt{D29\_P0}, this instance unveils the potential of \texttt{Q4RPD} to solve large and complex scenarios. In fact, this is the use case with the most paths to be calculated.
\end{itemize}

\begin{figure}[t]
    \centering
    \includegraphics[width=0.80\linewidth]{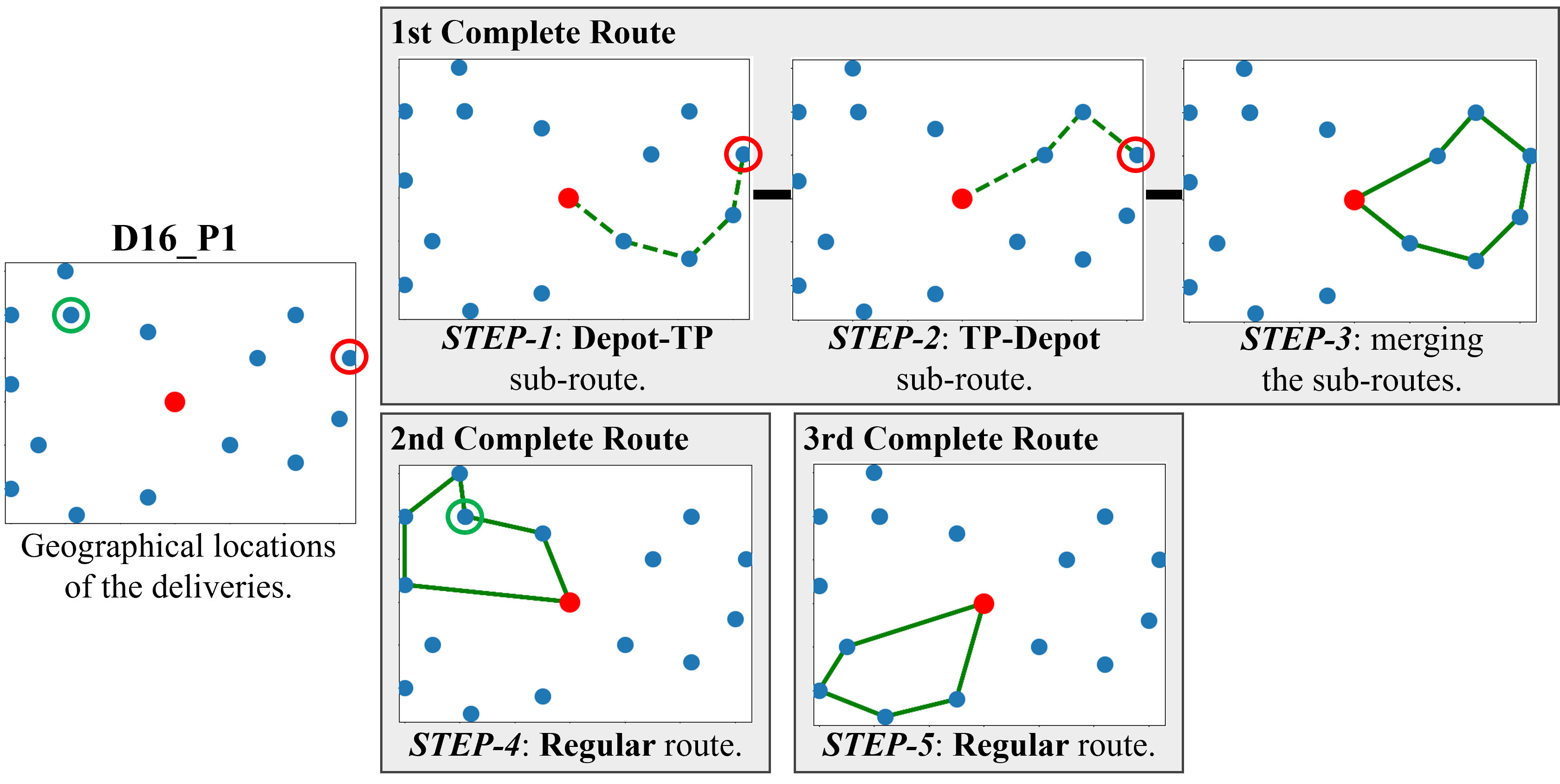}
    \caption{Step-by-step resolution of \texttt{D16\_P1}, consisting of non-priority 15 deliveries and one TP (surrounded by a red circle). Two non-priority demands belong to the same client (surrounded by a green circle), which are served by the same truck.}
    \label{fig:D17P1R3}
\end{figure}

\begin{figure}[t]
    \centering
    \includegraphics[width=0.8\linewidth]{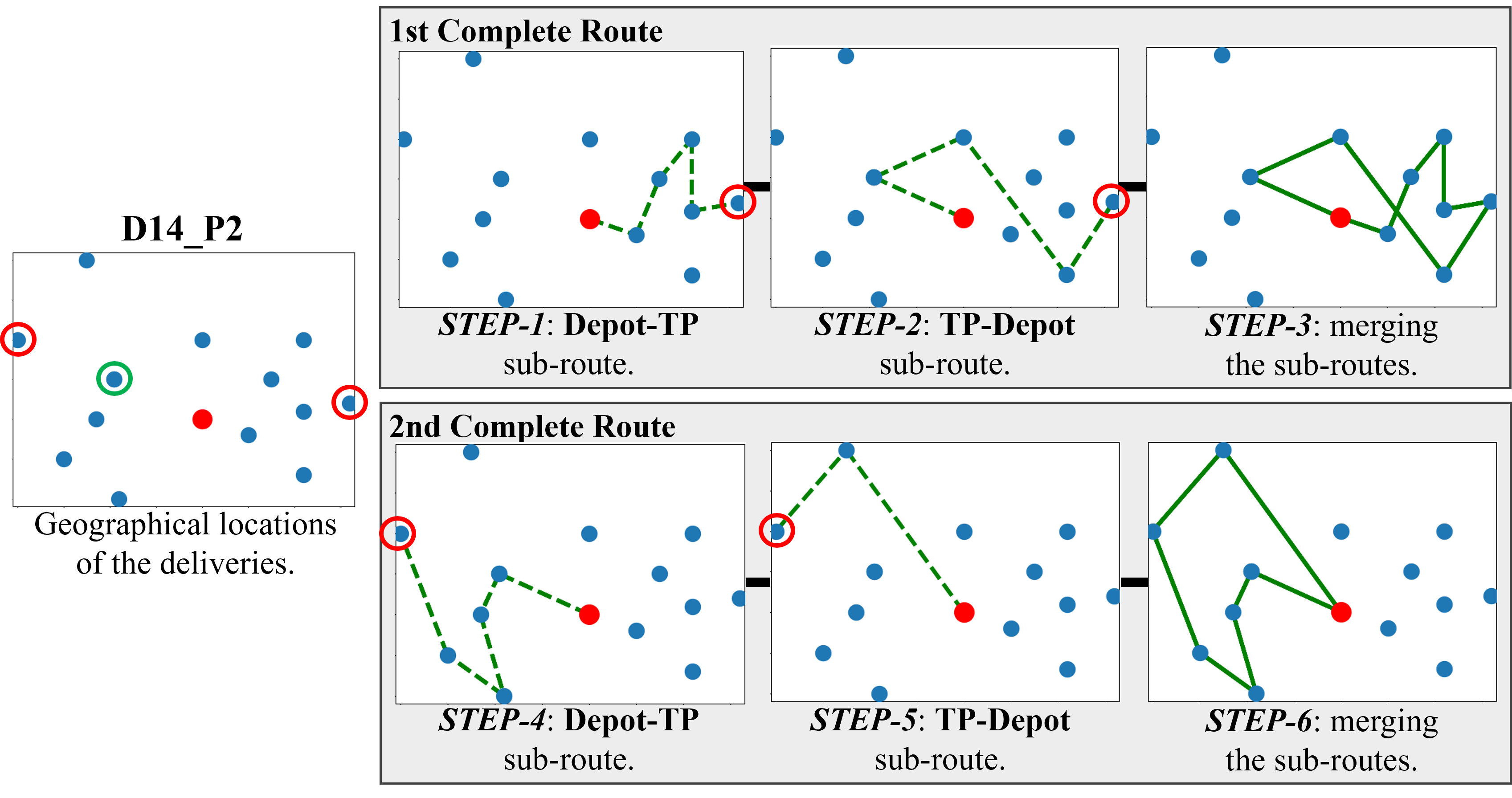}
    \caption{Step-by-step resolution of \texttt{D14\_P2}, composed of 12 non-priority deliveries and two TP (surrounded by a red circle). Also, two non-priority demands correspond to the same client (surrounded by a green circle), which must be served separately.}
    \label{fig:D15P2R2}
\end{figure}

\begin{figure}[t]
    \centering
    \includegraphics[width=1.0\linewidth]{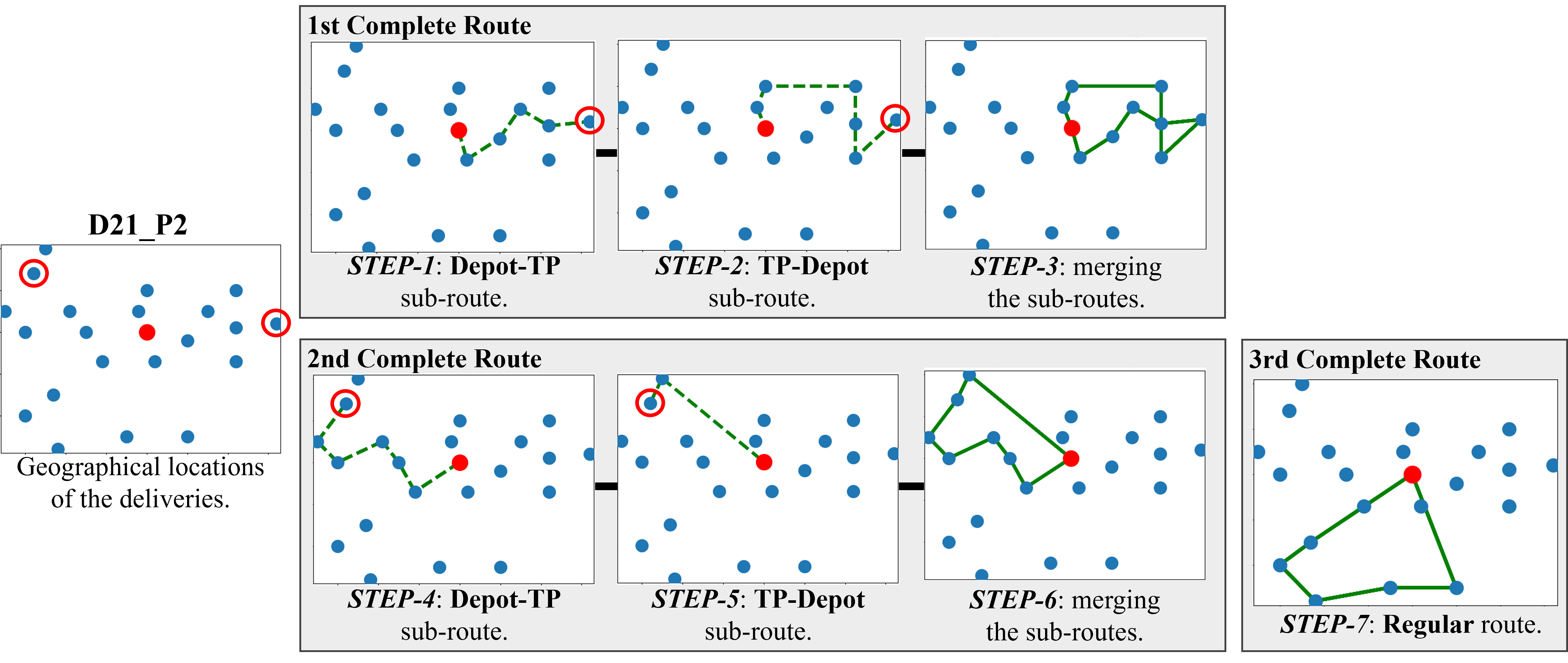}
    \caption{Step-by-step resolution of \texttt{D21\_P2}, with 19 non-priority deliveries and two TP (surrounded by a red circle).}
    \label{fig:D22P2R3}
\end{figure}

\starsubsection{Performance analysis}\label{sec:pa}
Concerning the \texttt{Q4RPD} performance, it is important to mention that stage \textit{\textbf{S1}} (described in \seclabel{sec:problem}) involved in the pre-selection of the (sub-)route and truck reduces the search space and therefore may affect the global performance of the solver. This means that by attending to the \textit{\textbf{business preferences}} indicated by Ertransit, the heuristics may hinder or even prevent the optimal solution from being reached. Nevertheless, this stage must be computed in advance to fix the upper bound parameters that complete the mathematical formulation that eventually sets in motion the (sub-)route construction. In addition, this classical processing forces the application of the preferences \textit{P1} (i.e., one truck for each complete route) and \textit{P2} (i.e., owned trucks have been prioritized to minimize the cost associated with the rental of trucks). Lastly, and after conducting this experimentation, we have found that \textit{P3} is not always covered by the currently proposed scheme. Even though this setback only occurs in rare cases, future versions of the scheme will introduce more intelligent resource allocation schemes to ensure the fulfillment of this preference. 

\begin{figure}[t]
    \centering
    \includegraphics[width=0.82\linewidth]{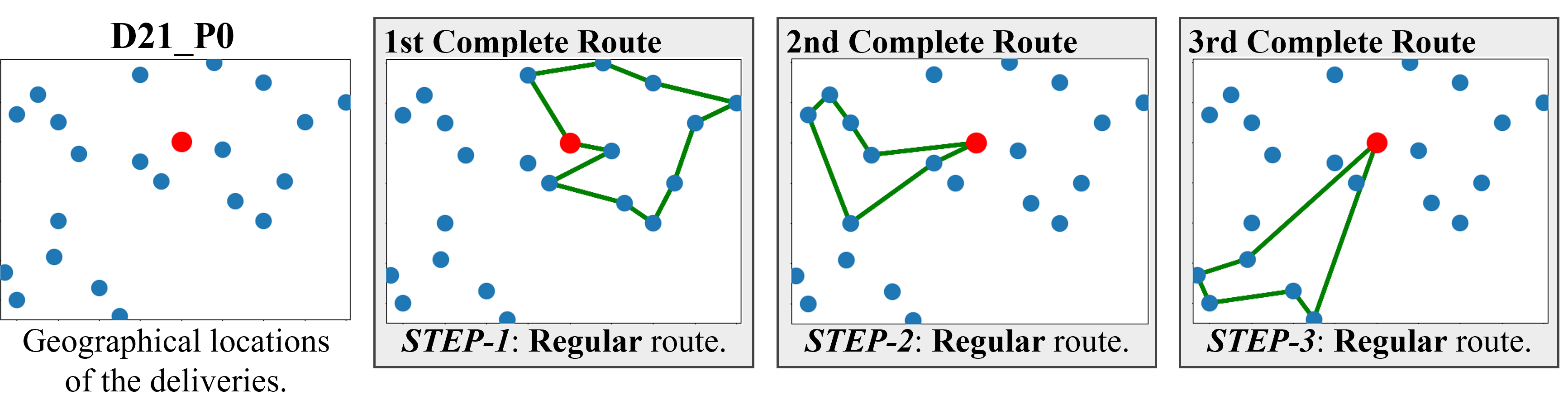}
    \caption{Step-by-step resolution of \texttt{D21\_P0}, composed of non-priority 21 deliveries.}
    \label{fig:D22P0R3}
\end{figure}

\begin{figure}[t]
    \centering
    \includegraphics[width=0.9\linewidth]{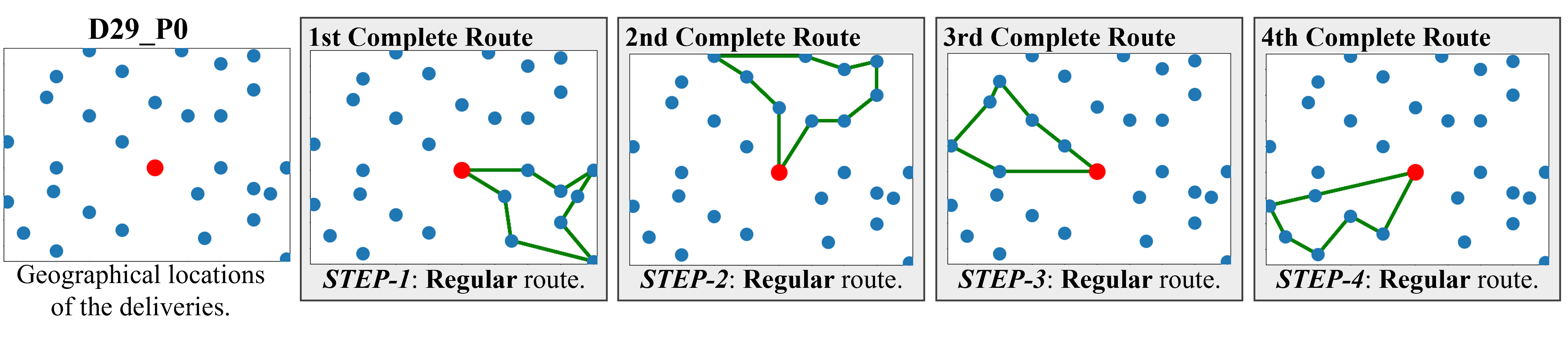}
    \caption{Step-by-step resolution of \texttt{D29\_P0}, composed of no-priority 29 deliveries.}
    \label{fig:D30P0R4}
\end{figure}

\noindent\begin{table}[!t]
 \centering
 \resizebox{0.70\columnwidth}{!}{
     \begin{tabular}{|l|l|r|r|r|r|r|r|c|}
      \hline
       \multicolumn{2}{|c|}{\multirow{2}{*}{\textbf{Instance}}} & \multicolumn{2}{c|}{\textbf{Problem solution}} & \multicolumn{3}{c|}{\textbf{Validation}} & \multicolumn{2}{c|}{\textbf{Objectives of the SRP}}\\
       
       \multicolumn{2}{|c|}{} & \#Full routes & \multicolumn{1}{c|}{\makecell[r]{\#(Sub-)routes}} & R1 & R2 & R3 & \multicolumn{1}{c|}{$\sum o_1$} & $o_2$ optimality\\ \hline
      \texttt{D14\_P1} & \figabvlabel{fig:D15P1R2} & 2 & [1,1,0,1] & $\checkmark$ & $\checkmark$ & $\checkmark$ & 210.43 (+ 4.3\%) & $\checkmark$\\ \hline
      
      \texttt{D16\_P1} & \figabvlabel{fig:D17P1R3} & 3 & [2,1,0,1] & $\checkmark$ & $\checkmark$ & $\checkmark$ & 223.74 (+0.0\%) & $\checkmark$\\  \hline
      
      \texttt{D14\_P2} & \figabvlabel{fig:D15P2R2} & 2 & [0,2,0,2] & $\checkmark$ & $\checkmark$ & $\checkmark$ & 245.30 (+ 6.3\%) & $\checkmark$\\ \hline 
      
      \texttt{D21\_P2} & \figabvlabel{fig:D22P2R3} & 3 & [1,2,0,2] & $\checkmark$ & $\checkmark$ & $\checkmark$ & 309.99 (+3.5\%) & $\checkmark$\\ \hline 
      
      \texttt{D21\_P0} & \figabvlabel{fig:D22P0R3} & 3 & [3,0,0,0] & $\checkmark$ & $-$ & $\checkmark$ & 381.46 (+0.0\%) & $\checkmark$ \\ \hline
      
      \texttt{D29\_P0} & \figabvlabel{fig:D30P0R4} & 4 & [4,0,0,0] & $\checkmark$ & $-$ & $\checkmark$ & 562.11 (+0.0\%) & $\checkmark$\\ \hline
     \end{tabular}
}

\caption{Summary of the results obtained by \texttt{Q4RPD}. Number of (sub-)routes is represented as [A,B,C,D] where A=\textit{regular} routes; B: \textit{Depot-\texttt{TP}} sub-routes; C: \textit{\texttt{TP}-\texttt{TP}} sub-routes and D: \textit{\texttt{TP}-Depot} sub-routes. In \textit{Validation} section:$\checkmark$=restriction fulfilled; $\times$=restriction broken; and $-$: constraint not present in the instance. In \textit{Objectives of the SRP} section: value in brackets represent the distance difference of the sums of $o_1$ when using our solver \texttt{Q4RPD} from the optimal result found by \texttt{Google OR-Tools}; while $\checkmark$ in last column represents the optimality in all the $o_2$ computed.}
\label{tab:results}
\end{table}%

Having said this, the \texttt{Q4RPD} performance of the \textit{\textbf{objective function}} and the \textit{\textbf{constraints}} for the six instances are summarized in Table \ref{tab:results}. Along with the reference to the figure in which the best-found solution of the instance is represented, the following information, divided into three main blocks, is provided in this table: 

\begin{itemize}
    \item \textit{Problem solutions}: the number of complete routes that compose the best solution found by \texttt{Q4RPD}; and the number and type of (sub-)routes of that solution.
    \item \textit{Validation}: in this second section, the fulfillment of constraints \textit{R1} (i.e., truck capacities), \textit{R2} (i.e., time constraints of \texttt{TP} deliveries), and \textit{R3} (i.e., the length of the working day) is assessed.
    \item \textit{Objectives of the \texttt{SRP}}: concerning the efficiency of the solutions found, and for comparison purposes, a classic TSP has been selected as the baseline algorithm to assess the performance of our solution. Concretely, each route built by \texttt{Q4RPD} is evaluated against a TSP solution obtained from \texttt{Google OR-Tools}, which has been used for this study because of the following reasons: \textit{i)} it is open-source software; \textit{ii)} it has been frequently employed for benchmarking purposes \cite{pan2023h,silva2023capacitated,guo2024imtsp}; and \textit{iii)} it has demonstrated a good performance for solving combinatorial optimization problems \cite{da2019google}. 
    
    It is important to note that \texttt{Google OR-Tools} only calculates the canonical TSP. Our solver \texttt{Q4RPD}, in turn, is forced to meet the \texttt{TP}s schedule imposed by \textit{R1}, which causes tweaks in the path that increase the distance but avoids violating this constraint. Specifically, the sum of all the $o_1$ objectives calculated by \texttt{Q4RPD} is displayed with a percentage representing the difference with respect to the sum of the costs of the solutions obtained by \texttt{Google OR-Tools}. In addition to that, we also represent in Table \ref{tab:results} if all $o_2$ calculated by \texttt{Q4RPD} have been optimized, i.e., if all the generated (sub-)routes maximize the number of deliveries attended while meeting restrictions \textit{R1} and \textit{R3}.
\end{itemize} 

Two main conclusions can be drawn from the results. First, \texttt{Q4RPD} has proved to be a promising solver for dealing with \texttt{2DH-PDP}. Regarding the performance of the solver in terms of constraint fulfillment and its capability to minimize the objective function:

\begin{itemize}
    \item \texttt{Q4RPD} has emerged as a suitable solver to efficiently deal with the constraints imposed on the problem, always providing solutions that meet the imposed restrictions.
    \item In terms of distance minimization, it can be seen that \texttt{Q4RPD} can compete with a classical solver such as \texttt{Google OR-Tools}. On the one hand, in cases where there are no \texttt{TP} deliveries, i.e., \texttt{D21\_P0} and \texttt{D29\_P0}, the routes calculated by \texttt{Q4RPD} have been demonstrated to be the optimal ones. On the other hand, in those instances with \texttt{TP} deliveries, \texttt{Q4RPD} provide solutions with a deviation of less than 6.3\% in all cases. It should be noted that this deviation in the objective $o_1$ is not attributed to the performance of the routing algorithm but to the obligation of \texttt{Q4RPD} to comply with the restriction \textit{R2}. In fact, in those cases in which \texttt{Google OR-Tools} provides better results in terms of $o_1$, the restriction \textit{R2} is not met. All this allows certifying the good performance of \texttt{Q4RPD} for the scenarios tackled in this research. 
\end{itemize}

Last but not least, the iterative nature of \texttt{Q4RPD} permits solving instances of a larger size than usually seen in the current NISQ-era literature. While \texttt{Q4RPD} can efficiently cope with problems such as \texttt{D29\_P0}, composed of 30 nodes, recent studies on routing problems through the QC perspective work with significantly smaller problems \cite{mori2023quantum,qian2023comparative,le2023quantum,delgado2022quantum,tambunan2022quantum,mohanty2023analysis}. Finally, for the sake of replicability, all the instances employed in this experimentation as well as the detailed outcomes are openly available\cite{PDPData}.

\starsection{Conclusions and future work}\label{sec:conc}

In this paper, a quantum-classical system for solving a real-world logistic problem is presented, named \textit{Quantum for Real Package Delivery} (\texttt{Q4RPD}). Specifically, this solving scheme deals with complex routing problems whose specifications include constraints such as a heterogeneous fleet of vehicles, priority deliveries, package description, and truck capacity according to weight and dimensions. For this reason, the problem has been coined as \textit{2-dimensional and Heterogeneous Package Delivery with Priorities} (\texttt{2DH-PDP}). To demonstrate the applicability, six instances of different sizes and characteristics have been utilized. Several inspiring challenges and opportunities have been identified for future work, graphically summarized and categorized in Fig.\ref{fig:futurework}: 
\begin{itemize}
    \item As for the \textbf{problem definition}:
    \begin{itemize}
        \item An extension of the mathematical formulation of \texttt{2DH-PDP} to include further real-world situations. Some examples of new features would add more realistic and/or complex cost calculations with the cost of the node-to-node trip subject to pre-calculated fuel consumption, the likelihood of road congestion, 
        \item The use of 3D dimensions, i.e., depth, width, and height, to categorize packages and truck capacities. This line would relate this work to other branches of research, such as the bin packing problem \cite{v2023hybrid}.
        \item The possibility of reusing the trucks to complete multiple routes.
    \end{itemize}
    \item As for the \textbf{future applications}:
    \begin{itemize}
        \item The extension of \texttt{Q4RPD} to companies with long-distance deliveries that would exceed the one-day delivery time. 
        \item Consider other transportation options as possible means of transport, such as cargo flights or ships.
    \end{itemize}

\begin{figure}[H]
    \centering
    \includegraphics[width=0.9\linewidth]{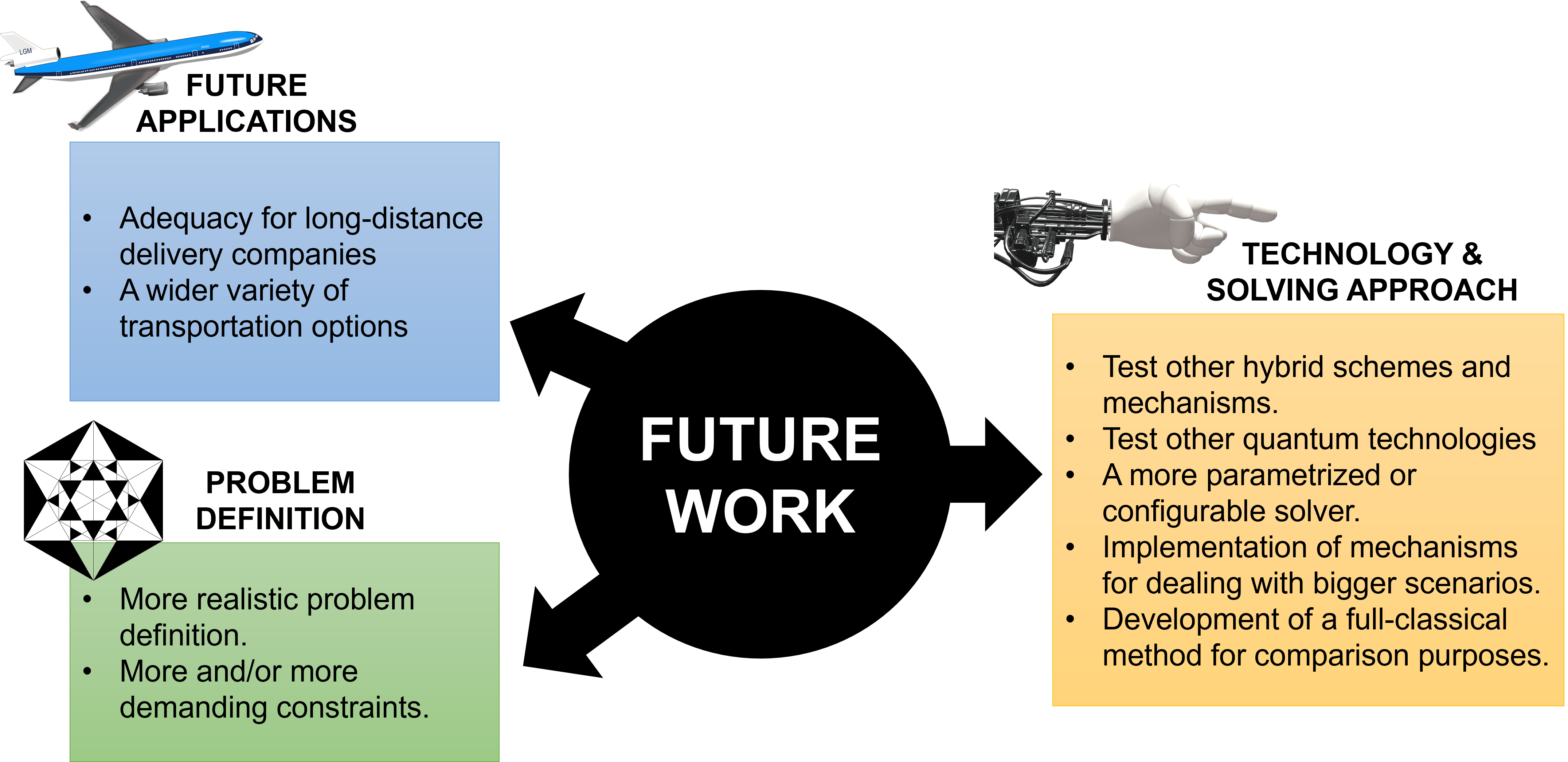}
    \caption{A graphical summary of the planned future work.}
    \label{fig:futurework}
\end{figure}

    \item With respect to \textbf{technology and the solving approach}:
    \begin{itemize}
        \item The use of complexity reduction mechanisms to deal with larger scenarios efficiently by reducing the problem's size while maintaining all the constraints of the problem.
        \item The design and implementation of hybrid methods incorporating another QC paradigm or technology. 
        \item The development of more configurable solvers that could automatically handle different company preferences or easily adapt to other companies' requirements. Another interesting avenue would be to grant the solver the ability to transform heuristics, i.e., preferences, into constraints or sub-objectives in the objective function, and the other way around.
        \item The implementation of a full-classical algorithm to make a performance comparison with \texttt{Q4RPD}.
    \end{itemize}    
\end{itemize}  

\starsection{Data availability}\label{sec:data}

The data used, as well as the results discussed, are available at \url{http://dx.doi.org/10.17632/yv48pwk96y.1}.

\bibliography{sample}

\starsection{Acknowledgements}

This work was supported by the Basque Government through HAZITEK program (Q4\_Real project, ZE-2022/00033) and through Plan complementario comunicación cuántica (EXP. 2022/01341) (A/20220551). The authors thank Miguel Angel Casla, Ertransit’s Director of Operations, for his assistance throughout this research.

\starsection{Author contributions statement}

All authors conceived the research and defined the problem. All authors conceived the experiments. E.O. and E.V.R design the solver. E.O. developed the code and conducted the experimentation. E.O. and E.V.R. wrote the manuscript. All authors reviewed the manuscript.

\starsection{Competing interests}

The authors declare no competing interests.

\end{document}